\documentclass[journal,draftclsnofoot,onecolumn,12pt,twoside]{IEEEtran}
\pdfoutput=1

\usepackage{cite}

\usepackage{graphicx}
\usepackage{epstopdf}
\usepackage{pdflscape}
\DeclareGraphicsExtensions{.eps}
\usepackage{array}
\usepackage{hhline}
\usepackage{multirow}
\usepackage{booktabs}
\usepackage{stfloats}
\usepackage{soul}
\usepackage[cmex10]{amsmath}
\usepackage{amsfonts,amsmath,amssymb,epsfig,epstopdf, color}

\newcommand{\norm}[1]{\left\lVert#1\right\rVert}

\usepackage[tight,footnotesize]{subfigure}

\usepackage{url}

\begin{document}

\title{Hybrid Analog-Digital Precoding \\ Revisited under Realistic RF Modeling}
\author{\IEEEauthorblockN{Adrian Garcia-Rodriguez,~\IEEEmembership{Student Member,~IEEE},  Vijay Venkateswaran,~\IEEEmembership{Member,~IEEE}, Pawel Rulikowski,~\IEEEmembership{Member,~IEEE}, \\ and Christos Masouros,~\IEEEmembership{Senior Member,~IEEE}}
\thanks{A.\ Garcia-Rodriguez and C.\ Masouros are with the Department of Electronic and Electrical Engineering, University College London, UK (e-mail: $\{$adrian.rodriguez.12, c.masouros$\}$@ucl.ac.uk). The work of A.\ Garcia-Rodriguez was carried out in Alcatel-Lucent Bell Labs, Ireland.}
\thanks{V.\ Venkateswaran is currently with Huawei Technologies, Sweden (e-mail: vijay.venkateswaran@huawei.com), and was with Alcatel-Lucent Bell Labs, Ireland, where the work was carried out.}
\thanks{P.\ Rulikowski is with Nokia Bell Labs, Ireland (e-mail: pawel.rulikowski@nokia.com).}
\thanks{This research was supported in part by the Royal Academy of Engineering, UK and the EPSRC under grant EP/M014150/1.}}

\maketitle

\vspace*{-1cm}
\begin{abstract}
In this paper we revisit hybrid analog-digital precoding systems with emphasis on their modelling and radio-frequency (RF) losses, to realistically evaluate their benefits in 5G system implementations. For this, we decompose the analog beamforming networks (ABFN) as a bank of commonly used RF components and formulate realistic model constraints based on their S-parameters. Specifically, we concentrate on fully-connected ABFN (FC-ABFN) and Butler networks for implementing the discrete Fourier transform (DFT) in the RF domain. The results presented in this paper reveal that the performance and energy efficiency of hybrid precoding systems are severely affected, once practical factors are considered in the overall design. In this context, we also show that Butler RF networks are capable of providing better performances than FC-ABFN for systems with a large number of RF chains.
\end{abstract}

\begin{keywords} \noindent Analog beamforming networks, hybrid precoding, millimeter wave, massive MIMO, Butler matrix.
\end{keywords}

\IEEEpeerreviewmaketitle 

\section{Introduction}

5G communication systems are expected to incorporate a large number of antennas at the base stations (BS) for serving a multiplicity of user terminals while satisfying their data rate requirements \cite{6375940,6736750}. For instance, both millimeter-wave (mmWave) and large-scale antenna systems (LSAS) exploit the large number of antennas available for compensating the severe path loss at high frequencies and providing a favourable propagation respectively \cite{6375940,6736750}. In turn, the substantial increase in the number of antennas has motivated the development of strategies with the essential objective of reducing both their hardware and signal processing complexity challenges involved in \mbox{the design of future communications systems.}

Hybrid analog-digital precoding and detection schemes aim at reducing the number of radio-frequency (RF) chains  by translating part of the signal processing operations to the RF domain \cite{6736750}. Indeed, this approach is crucial in mmWave systems due to the reduced number of degrees of freedom offered by the communication channel and the need for providing beamforming gains \cite{6717211}. Moreover, reducing the number of RF chains and the digital processing load leads to EE improvements. In this context, a variety of ABFN have been recently proposed  \cite{6736750,6717211,6542746}. However, they generally disregard the practical implications of signal processing in the RF domain, which have been partly considered in \cite{1389158,DBLP:journals/corr/VenkateswaranPG15}.

In this letter we characterize the impact of considering realistic ABFN in the performance of hybrid precoding schemes. In particular, our contributions can be summarized as follows:
\begin{itemize}
\item We model ABFN as a bank of elementary RF components. The S-parameter representation of the FC-ABFN reveals that there are significant power losses even when ideal components are considered  -- a feature commonly ignored in the related literature and that promotes the implementation of alternative ABFN \cite{pozar2009microwave,DBLP:journals/corr/VenkateswaranPG15}.
\item Subsequently, we incorporate the insertion losses (IL) found in real analog hardware components for assessing the effective performance of hybrid precoding systems.
\end{itemize}

\section{System Model: Hybrid Precoding Systems}

\begin{figure}[!t]
	\centering
		\includegraphics[width=0.75\textwidth]{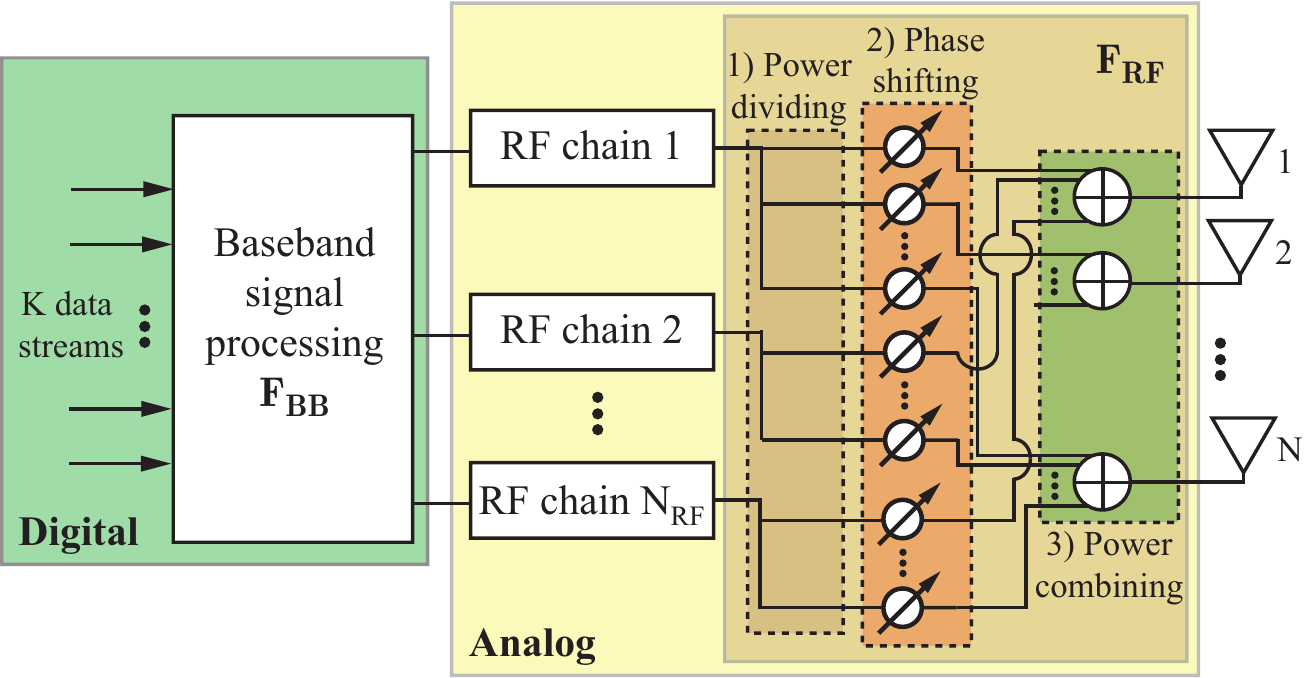}
		\caption{Block diagram of a hybrid precoding system comprised of digital precoding and a fully-connected analog beamforming network (FC-ABFN).}
	\label{fig:hybridPrecoding}
\end{figure}

Let us consider a base station (BS) comprised of $N$ antennas transmitting towards $K \leq N$ single-antenna users. This system can be characterized as 
\begin{equation}
\mathbf{y} = \mathbf{H}^{H} \mathbf{x} + \mathbf{w},
\end{equation}
where $\mathbf{x} \in \mathbb{C}^{N \times 1}$ and $\mathbf{y} \in \mathbb{C}^{K \times 1}$ denote the transmitted and received signals respectively, whereas $\mathbf{w}\in\mathbb{C}^{K \times1}\sim\mathcal{C}\mathcal{N}(\mathbf{0},\sigma^2\textbf{I}_{K})$ is the circularly symmetric additive white-Gaussian noise vector. Moreover, $\mathbf{h}^{H}_{k}$ collects the frequency-flat channel gains between the BS antennas and the $k$-th user, where $\mathbf{h}_{k} = \mathbf{R}_{k}^{\frac{1}{2}} \mathbf{z}_{k}$ denotes the $k$-th column of $\mathbf{H} \in \mathbb{C}^{N \times K}$ \cite{6542746}. Here, $\mathbf{z}_{k} \in\mathbb{C}^{N \times1}\sim\mathcal{C}\mathcal{N}(\mathbf{0},\textbf{I}_{N})$ and $\mathbf{R}_{k} \in \mathbb{C}^{N \times N}$ represents the channel covariance matrix. The transmitted signal in hybrid analog-digital precoding systems can be decomposed as \cite{6717211}
\begin{equation}
\mathbf{x} = \mathbf{F} \mathbf{s} =  \mathbf{F}_{\text{RF}} \mathbf{F}_{\text{BB}} \mathbf{s},
\end{equation}
where $\mathbf{s} \in \mathbb{C}^{K \times 1} \sim\mathcal{C}\mathcal{N}(\mathbf{0},\frac{1}{K}\textbf{I}_{K})$ comprises the modulated data symbols and $\mathbf{F} \in \mathbb{C}^{N \times K}$ is the composite precoding matrix. Additionally, $\mathbf{F}_{\text{BB}} \in \mathbb{C}^{N_{\text{RF}} \times K}$ represents the digital baseband precoding matrix and $\mathbf{F}_{\text{RF}} \in \mathbb{C}^{N \times N_{\text{RF}}}$ characterizes the ABFN. Here, $N_{\text{RF}} \geq K$ denotes the number of RF chains employed for transmission. An illustrative example of a hybrid precoding system is shown in \figurename~\ref{fig:hybridPrecoding}. The signal-to-interference-plus-noise ratio (SINR) of the $k$-th user is given by \cite{6172680}
\begin{equation}
\gamma_{k} = \frac{\vert \mathbf{h}^{H}_{k} \mathbf{f}_k \vert^{2}}{\sum_{i \neq k} \vert \mathbf{h}^{H}_{k} \mathbf{f}_i \vert^{2} + \sigma^2}, 
\end{equation}
where $\mathbf{f}_k \in \mathbb{C}^{N \times 1}$ represents the $k$-th column of $\mathbf{F}$. The ergodic sum rates in bits  per second can be expressed as \cite{6172680}
\begin{equation}
R_{\text{sum}} = B S_{\text{e}} = B \sum_{k=1}^{K} \mathbb{E} \left[ \log_{2} \left( 1 + \gamma_{k} \right) \right],
\end{equation}
where $B$ denotes the system bandwidth and $S_{\text{e}}$ represents the sum spectral efficiency (SE). 

At this point we note that traditional hybrid system models normalize the composite precoding matrix $\mathbf{F}$ for satisfying a specific sum power constraint and subsequently apply additional constraints to the RF precoder $\mathbf{F}_{\text{RF}}$. Instead, here we aim at deriving the constraints of the RF precoder $\mathbf{F}_{\text{RF}}$ based on a S-parameter analysis of the specific hardware implementation of the ABFN. Specifically, we solely impose $\norm{\mathbf{F}_{\text{BB}}}_{F}^{2} = K$ and define $\rho \triangleq \frac{K}{\sigma^{2}}$. The above constraints guarantee a fair comparison between hybrid and fully digital systems ($\mathbf{F}_{\text{RF}} = \mathbf{I}_{N}$).

\section{Analog Beamforming Networks (ABFN)}
\label{sec:analogBFNetworks}

In this section we derive a realistic model for characterizing two popular ABFN architectures for the design of hybrid communication systems: Fully-connected and DFT-based ABFN. Our models are based on the detailed electromagnetic study of hybrid architectures implemented in \cite{DBLP:journals/corr/VenkateswaranPG15} and provide a realistic framework for obtaining a fair comparison between fully-digital solutions and practical hybrid precoding designs.

\subsection{Fully-Connected Analog Beamforming Network}
\label{sec:fullyConnectedABFN}

The architecture of a FC-ABFN is shown in \figurename~\ref{fig:hybridPrecoding}, where three stages can be clearly identified: a first one comprised of power dividers where each of the $N_{\text{RF}}$ input signals is divided into $N$ equal-power outputs characterized by the matrix $\mathbf{F}_{\text{D}} \in \mathbb{C}^{\left( N_{\text{RF}} \cdot N\right) \times N_{\text{RF}}}$, a subsequent one where $N_{\text{RF}} \cdot N$ signals are phase shifted represented by $\mathbf{F}_{\text{PS}} \in \mathbb{C}^{\left( N_{\text{RF}} \cdot N\right) \times \left( N_{\text{RF}} \cdot N\right)}$, and a final stage where $N_{\text{RF}}$ signals are combined with power combiners and coupled to $N$ antenna ports characterized by $\mathbf{F}_{\text{C}} \in \mathbb{C}^{N \times \left( N_{\text{RF}} \cdot N\right)}$. Based on the above and in order to offer a complete view of the ABFN's behaviour, we decompose the analog beamforming matrix as \cite{DBLP:journals/corr/VenkateswaranPG15}
\begin{equation}
\mathbf{F}_{\text{RF}} = \mathbf{F}_{\text{C}} \cdot \mathbf{F}_{\text{PS}} \cdot \mathbf{F}_{\text{D}}.
\label{eq:fullyConnectedAnalog}
\end{equation}
At this point we note that $\mathbf{F}_{\text{RF}}$ is inherently defined by $\mathbf{F}_{\text{PS}}$, since both $\mathbf{F}_{\text{C}}$ and $\mathbf{F}_{\text{D}}$ are fixed as shown in the following. A common design criterion for $\mathbf{F}_{\text{PS}}$ consists in selecting phase shifting values according to the transmit array response vectors at the angles of departure from the transmitter \cite{6717211,DBLP:journals/corr/VenkateswaranPG15}. However, in general the specific phase shifting values $\mathbf{F}_{\text{PS}}$ can be obtained following multiple design criteria \cite{6736750,6717211,6851941,6542746}. However, in general the specific phase shifting values $\mathbf{F}_{\text{PS}}$ can be obtained following multiple design criteria whose exhaustive description is out of the scope of this paper, since they do not modify the conclusions derived in the \mbox{following \cite{6736750,6717211,6851941,6542746}.}

Note that the decomposition in \eqref{eq:fullyConnectedAnalog} is performed in the RF domain. Therefore, an accurate description of their operation should be based on understanding the RF characteristics of the specific components. For this reason we define the signal distribution in $\mathbf{F}_{\text{RF}}$ based on the S-parameter representation of the hardware components involved in $\mathbf{F}_{\text{C}}$,  $ \mathbf{F}_{\text{PS}}$ and $\mathbf{F}_{\text{D}}$. Specifically, $\mathbf{F}_{\text{D}}$, which is comprised of Wilkinson power dividers \cite{pozar2009microwave}, can be \mbox{modeled following a block diagonal structure}
\begin{equation}
\mathbf{F}_{\text{D}} =\sqrt{\frac{1}{L_{\text{S}} N}}\left( \begin{array}{cccc}
\mathbf{1}_{N} & \mathbf{0}_{N} & \ldots  & \mathbf{0}_{N}\\
\mathbf{0}_{N} & \mathbf{1}_{N} &  \ldots & \mathbf{0}_{N}\\
\vdots & \vdots &  \ddots   &\vdots \\
\mathbf{0}_{N}     & \mathbf{0}_{N} &  \ldots & \mathbf{1}_{N}  \end{array} \right),
\end{equation}
where $L_{\text{S}}$ corresponds to the substrate or \emph{static}  power loss \cite{pozar2009microwave}, and $\mathbf{1}_{T} \in \mathbb{N}^{T \times 1}$ and $\mathbf{0}_{T} \in \mathbb{N}^{T \times 1}$ represent the all-ones and all-zeros vectors respectively. The phase shifting network matrix $\mathbf{F}_{\text{PS}}$ is a diagonal matrix characterized by
\begin{equation}
\mathbf{F}_{\text{PS}} = \sqrt{1/L_{\text{PS}}} \cdot \text{diag} \left( [ f_{1,1}, f_{2,1}, \ldots , f_{N, N_{\text{RF}}}] \right),
\end{equation}
where $L_{\text{PS}}$ denotes the static power losses introduced by each phase shifter, and the coefficients $f_{i,j}, \forall \mbox{ } i \in \{ 1, \ldots, N\} \text{ and } j \in \{ 1, \ldots, N_{\text{RF}} \}$ corresponds to the $i,j$-th phase shift of $\mathbf{F}_{\text{RF}}$ normalized to satisfy $\norm{f_{i,j}} = 1$. The combining matrix $\mathbf{F}_{\text{C}}$ can be expressed as
\begin{equation}
\mathbf{F}_{\text{C}} =\sqrt{\frac{1}{L_{\text{C}}  N_{\text{RF}}}} \left[ \text{diag} \left( \mathbf{1}_{N} \right), \ldots,  \text{diag} \left( \mathbf{1}_{N} \right) \right].
\label{eq:combiningMatrix}
\end{equation}
With respect to the losses in the combining stage, there are two dominant factors: First, $L_{\text{C}}$ represents the static power losses introduced by the power combiners. Secondly, the S-parameter representation of passive RF components reveals additional losses in the form of the scaling coefficient $1/\sqrt{N_{\text{RF}}}$ in \eqref{eq:combiningMatrix}. In other words, the adaptive nature of $\mathbf{F}_{\text{BB}}$ and the data symbols produce phase and amplitude mismatches in the signals at the input of the power combiners, hence introducing a loss in the signal combining process - an aspect not often considered in the related literature. We refer to this loss as \textit{dynamic} power loss and we remark that it arises even for lossless (ideal) analog hardware components \cite{pozar2009microwave}. The consideration of the dynamic power losses entails that, in contrast with fully-digital precoding,  the power amplifiers will have to compensate for substantial signal-dependent losses in order to guarantee a given transmission power. Indeed, \eqref{eq:combiningMatrix} manifest power losses that scale linearly with $N_{\text{RF}}$  in ideal FC-ABFN.

\subsection{DFT Analog Beamforming Networks via Butler Matrices}
\label{sec:DFTButler}

\begin{figure}[!t]
	\centering
		\includegraphics[width=0.6\textwidth]{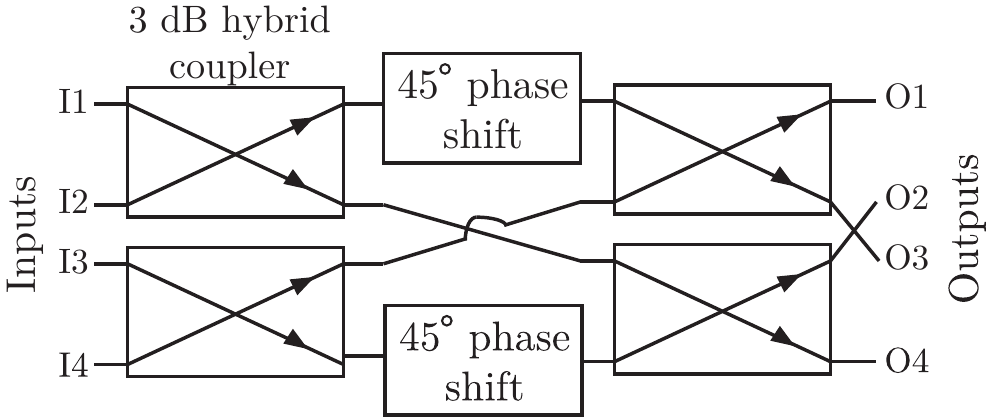}
		\caption{Block diagram of a $4 \times 4$ Butler matrix.}
	\label{fig:butlerMatrix}
\end{figure}

While fully-connected networks allow designing arbitrary ABFN, the above discussion has revealed that power combiners introduce substantial power losses. To alleviate these losses, we propose to consider $4$-port hybrid directional couplers instead of power combiners \cite{pozar2009microwave}. $4$-port hybrid couplers can be seen as a variation of $2$-point DFT, hence enabling the implementation of higher order DFTs in the RF domain by stacking multiple hybrid couplers and phase shifters consecutively \cite{4642967,6542746}. Indeed a variation of this approach, commonly referred to as Butler matrix, has been employed to generate orthogonal beams with minimal loss \cite{4642967,6542746}. An illustrative example of a $4 \times 4$ Butler matrix is shown in \figurename~\ref{fig:butlerMatrix}, where the conventional structure comprised of $N_{\text{HYB}} = \log_{2} \left( N \right)$ and $N_{\text{PS}} = \log_{2} \left( N \right) - 1 \mbox{ } (N \geq 2)$ hybrid coupling and phase shifting subsequent stages can be observed respectively. Based on the above, $\mathbf{F}_{\text{RF}}$ can be expressed in the case of Butler matrices as
\begin{equation}
\mathbf{F}_{\text{RF}} = \frac{1}{\sqrt{\left(L_{\text{PS}} \right)^{N_{\text{PS}}} \left(L_{\text{HYB}} \right)^{N_{\text{HYB}}} }} \mathbf{E}_{N_{\text{RF}}},
\label{eq:Butler}
\end{equation}
where $L_{\text{HYB}}$ are the static power losses introduced by each hybrid power coupler and $\mathbf{E}_{N_{\text{RF}}} \in \mathbb{C}^{N \times N_{\text{RF}}}$ is a submatrix of the $\mathbb{C}^{N \times N}$ DFT matrix \cite{6542746}. For uniform linear arrays (ULA), $\mathbf{E}_{N_{\text{RF}}}$ is defined by approximating $\mathbf{R}_{k}$ by a circulant matrix $\mathbf{C}_{k}$ and selecting the eigenvectors corresponding to its largest eigenvalues, since the eigenvectors of $\mathbf{C}_{k}$ \mbox{form a DFT matrix \cite{6542746}.}

\subsection{Static Insertion Losses (IL)}

The IL introduced by the additional RF components employed for hybrid beamforming should be incorporated into a realistic system model for deriving the additional power gains required in the RF stage and preserving the same output power of a fully digital precoding solution. Illustrative values of these signal-independent losses are shown in Table \ref{tab:lossesTable}. We highlight that, in general, the IL introduced in the mmWave band grow with the frequency of operation and that the values of Table \ref{tab:lossesTable} correspond to the Ka frequency band.


In the following we consider for simplicity that the $\left( T+1 \right)$-port power combiners and dividers required in large ABFN are built by concatenating $\log_{2} \left( T \right)$ three-port structures \cite{pozar2009microwave}. Therefore, the overall static losses for the splitting and combining stages are given by $L_{\text{S,dB}} = \bar{L}_{\text{S}} \log_{2}  \left( N \right)$ and $L_{\text{C,dB}} = \bar{L}_{\text{C}} \log_{2} \left( N_{\text{RF}} \right)$, where both $\bar{L}_{\text{S}}$ and $\bar{L}_{\text{C}}$ are provided in \mbox{Table \ref{tab:lossesTable}.} Indeed, the above power loss characterizations have been verified via the RF simulation of a $32 \times 32$ Butler matrix in Keysight's Advance Design System (ADS) \cite{ADSRef} using micro-strip lines on a Rogers $4350$ substrate material with dielectric constant $3.48$ and loss tangent $0.004$. For an illustrative frequency of $f = 2.6$ GHz and arbitrary $\mathbf{F}_{\text{BB}}$, we have observed that the dynamic loss is approximately zero, as described in Sec.\ \ref{sec:DFTButler}, and that \mbox{the static loss approaches $2.8$ dB.}

\begin{table}[!t] 
\begin{center}
{\renewcommand{\arraystretch}{0.9}
\caption{Orientative insertion losses (IL) of the hardware components employed in the design of analog beamforming networks.}
\begin{tabular}{| >{\centering\arraybackslash}p{6.5cm} ||  >{\centering\arraybackslash}p{4cm} | >{\centering\arraybackslash}p{4cm}|}
\hline
- & Millimeter-wave & Sub 5 GHz \\ \hline  \hline
Three-port power dividers / combiners $\left( \bar{L}_{\left\{ \text{S,C} \right\}} \right)$ & 0.6 dB \cite{Quinstar:powerdiv} & 0.5 dB \cite{Anaren:powerdiv}\\
\hline
Hybrid couplers & 0.5 dB \cite{Millitech:hybrid} & 0.15 dB \cite{Anaren:hybrid}\\
\hline
Phase shifters & 0.5 dB \cite{Millitech:phase} & 3.5 dB \cite{Hittite:mmic} \\
\hline
\end{tabular} 
\label{tab:lossesTable}}
\end{center}
\end{table}

\section{Energy Efficiency}
\label{sec:EE}

While reducing the number of RF transceivers, hybrid precoding schemes simultaneously incur in additional power losses as detailed in Sec.\ \ref{sec:analogBFNetworks}. A relevant reason for reducing the number of active RF chains is enhancing the transmission's energy efficiency (EE), which is defined as \cite{7031971}
\begin{equation}
\epsilon = \frac{R_{\text{sum}}}{P_{\text{tot}}} = \frac{B \sum_{k=1}^{K} \mathbb{E} \left[ \log_{2} \left( 1 + \gamma_{k} \right) \right]}{P_{\text{PA}} + N_{\text{RF}} P_{\text{RF}}  + P_{\text{syn}}} \text{ bits/Joule},
\label{eq:energyEfficiencyMetric}
\end{equation}
where $P_{\text{tot}}$ expressed in Watts (W) refers to the total power employed for transmission and $P_{\text{PA}} = P_{\text{out}} / \eta$ denotes the power consumed by a power amplifier with efficiency $\eta = 0.39$ to produce a signal output power of $P_{\text{out}} = 40$ W \cite{7031971}. Note that the effective output power of realistic hybrid precoding schemes will be reduced when compared with their fully-digital counterparts due to both dynamic and static power losses. Moreover, $P_{\text{RF}} = 1$ W and $P_{\text{syn}} = 2$ W denote the power consumed by each RF chain and the frequency synthesizer respectively, derived by \mbox{elaborating the values in \cite{7031971}.}

\section{Numerical Results}
\label{sec:simulationResults}

In this section we characterize the performance and EE of realistic ABFN. Although the conclusions derived in the following can be applicable to a vast number of hybrid precoding designs, in the following we concentrate on the hybrid precoding scheme referred to as joint spatial division and multiplexing (JSDM) \cite{6542746}, since it admits both fully-connected and DFT-based designs. We assume a ULA and adopt the one-ring channel correlation model for microwave frequencies, where the $i,j$-th entry of $\mathbf{R}_{k}$ is given by \cite{6542746}
\begin{equation}
\{\mathbf{R}_{k}\}_{i,j}  = \frac{1}{2 \Delta} \int_{-\Delta}^{\Delta} e^{j \frac{2 \pi}{\lambda} d (i-j) \cos (\vartheta + \theta)} d \vartheta,
\label{eq:oneRingChannelModel}
\end{equation}
where $d$, $\Delta$ and $\lambda$ denote the inter-antenna spacing, angular spread and the wavelength respectively. Moreover, $\theta$ represents the central azimuth angle between the BS and the users. Similarly to \cite{6542746}, we let $d=0.5\lambda$, $\Delta = 15^{\circ}$ and consider three user groups comprised of $K_{g} = 4$ single-antenna users each with $\theta_{1} = -45^{\circ}$, $\theta_{2} = 0^{\circ}$ and $\theta_{3} = 45^{\circ}$. We implement zero-forcing (ZF) precoding with perfect channel state information in the digital domain and define $b_{g}$ as the number of RF chains in the ABFN dedicated to serve the users in group $g$, which depends on the eigenvalues of $\mathbf{R}_{k}$ \cite{6542746}. Without loss of generality, we consider asymmetric power ratios at the output of the power dividers and a sub 5 GHz transmission in this section, since the general conclusions and observations derived in this work are independent of the operating frequency.

\begin{figure}[!t]
	\centering
		\includegraphics[width=0.65\textwidth]{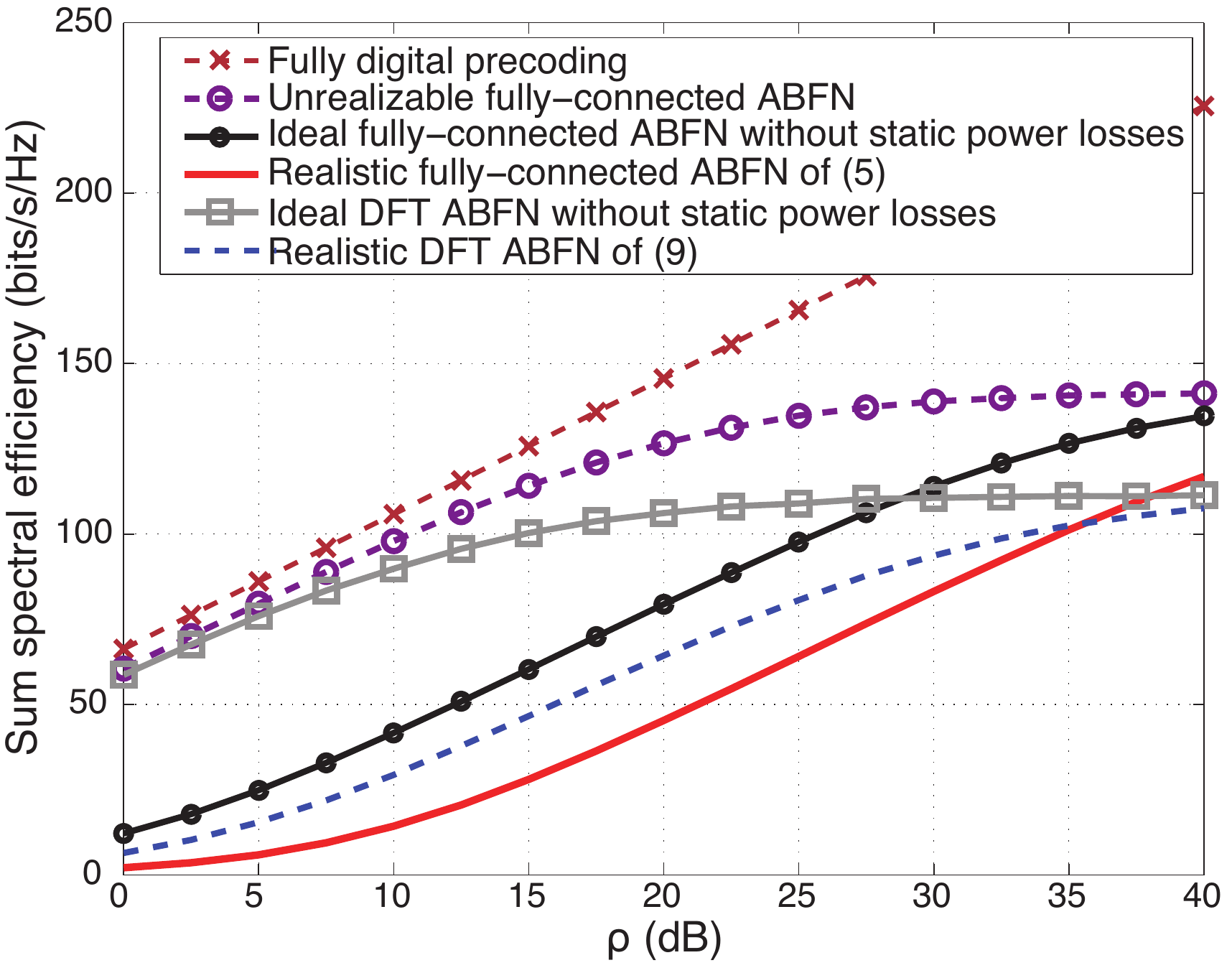}
		\caption{Sum spectral efficiency (bits/s/Hz) vs.\ $\rho$. $N = 64$, $K_{\left\{ 1,2,3 \right\} } = 4$, $b_{\left\{ 1,3 \right\}} = 10$ and $b_{2} = 12$.}
	\label{fig:SEvsRhoFull}
\end{figure}

\figurename~\ref{fig:SEvsRhoFull} considers $N = 64$ and shows the sum spectral efficiency (SE) against increasing $\rho$ for a fully digital precoding system and a hybrid JSDM system implemented via both DFT and FC-ABFN  ($N_{\text{RF}} = 32$). The results depicted in \figurename~\ref{fig:SEvsRhoFull} characterize the performance loss experienced by the realistic FC-ABFN even when ideal analog hardware components are considered, which can be explained by the dynamic power losses introduced by the signal combiners. The performance degradation becomes even more pronounced when the static IL are considered, making a realistic DFT network outperform the FC-ABFN for a large range of $\rho$ thanks to its reduced hardware losses. In this context, the results of \figurename~\ref{fig:SEvsRhoFull} also allow concluding  that hybrid coupler based DFT ABFN designs will be more spectrally efficient than FC-ABFN for large $N_{\text{RF}}$. This is because, as detailed in Sec.\ \ref{sec:fullyConnectedABFN}, the latter architecture introduces power losses that scale proportionally to $N_{\text{RF}}$ even when ideal analog hardware components are considered.

\begin{figure}[!t]
	\centering
		\includegraphics[width=0.65\textwidth]{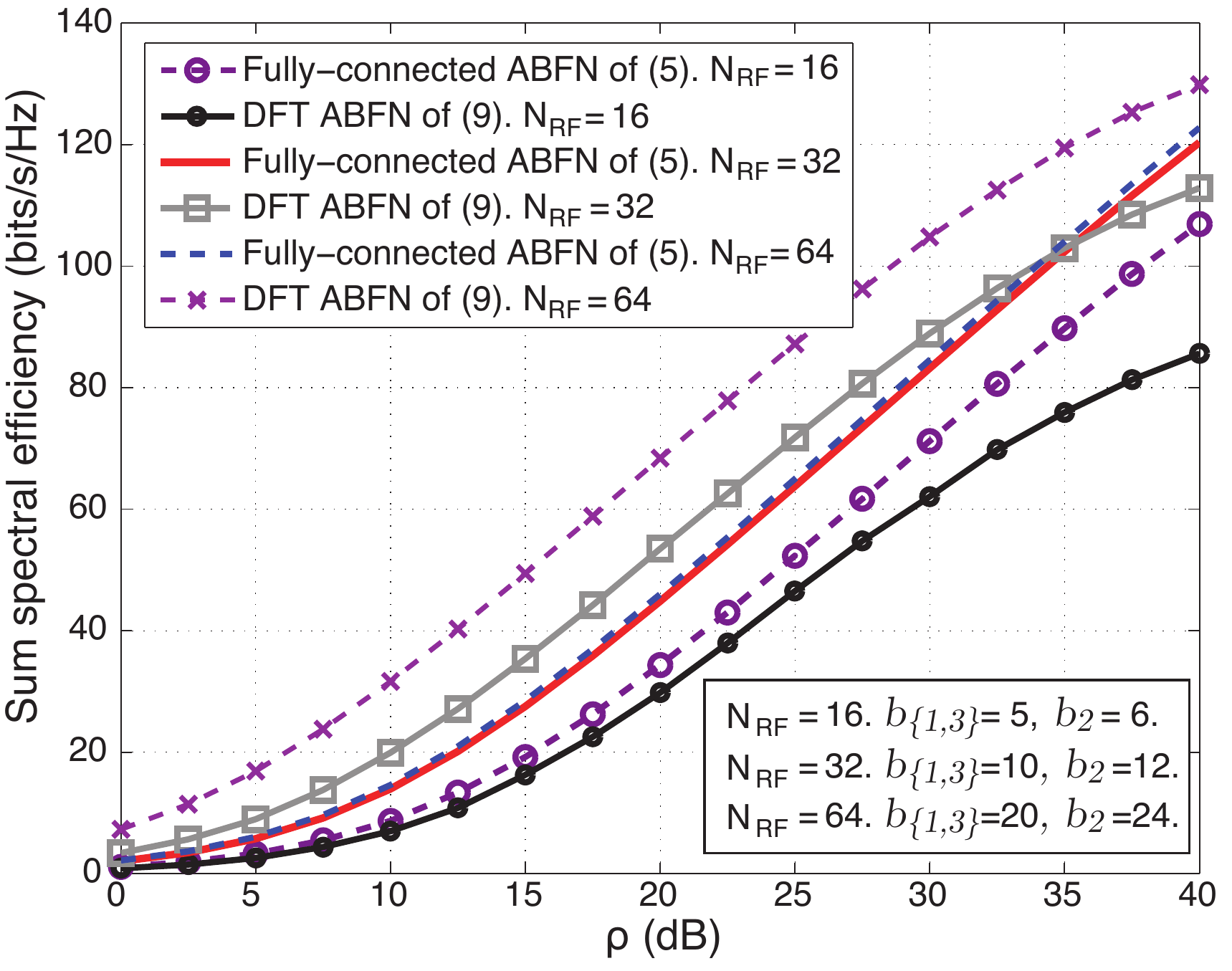}
		\caption{Sum spectral efficiency (bits/s/Hz) vs.\ $\rho$. $N = 128$, $K_{\left\{ 1,2,3 \right\} } = 4$ and varying number of RF chains.}
	\label{fig:SEvsRhoRFChains}
\end{figure}

\figurename~\ref{fig:SEvsRhoRFChains} shows the sum SE of realistic hybrid precoding schemes against $\rho$ in a system with $N = 128$ and different $N_{\text{RF}}$. It can be observed that the relative performance between DFT-based designs and FC-ABFN depends on both $N_{\text{RF}}$ and $\rho$. Moreover, \figurename~\ref{fig:SEvsRhoRFChains} shows that a sensible selection of $N_{\text{RF}}$ should consider the power losses' impact on the performance. For instance, it can be seen that while $N_{\text{RF}} = 64$ offers significant performance benefits over other alternatives for the DFT-based implementations, both $N_{\text{RF}} = 32$ and $N_{\text{RF}} = 64$ offer a similar performance for the case of FC-ABFN. This can be explained by noting that while implementing a larger $N_{\text{RF}}$ allows for improved flexibility in the hybrid design \cite{6717211}, the network losses of FC-ABFN also grow with $N_{\text{RF}}$ as per \eqref{eq:combiningMatrix}, something not quantified in the related literature.

\begin{figure}[!t]
	\centering
		\includegraphics[width=0.65\textwidth]{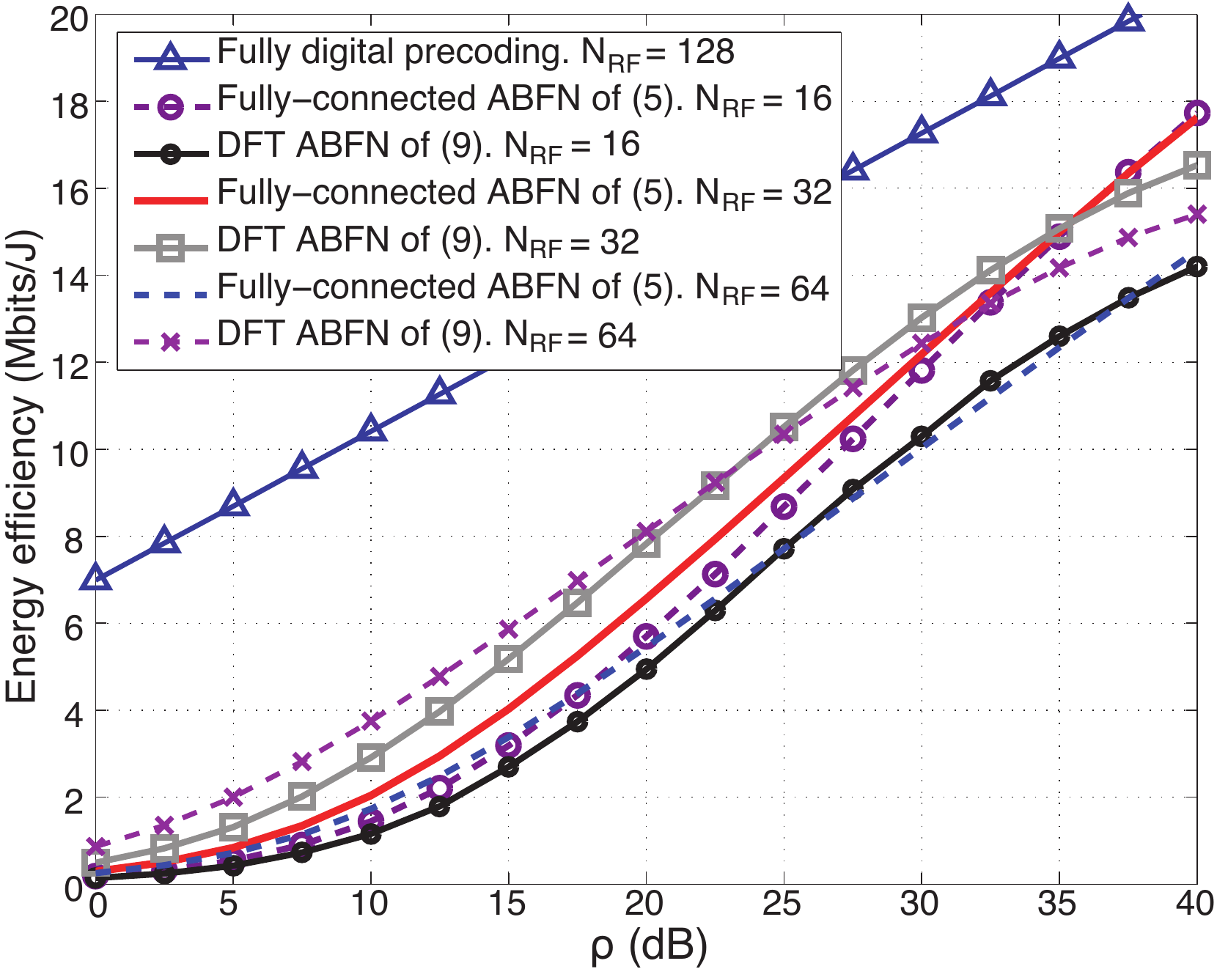}
		\caption{Energy efficiency (Mbits/Joule) vs.\ $\rho$. $N = 128$, $K_{\left\{ 1,2,3 \right\} } = 4$ and varying number of RF chains. $B = 20$ MHz and  $P_{\text{out}} = 46$ dBm.}
	\label{fig:EEvsRhoRFChains}
\end{figure}

\figurename~\ref{fig:EEvsRhoRFChains} represents the EE of the systems considered in \figurename~\ref{fig:SEvsRhoRFChains} with $B = 20$ MHz. The EE trends allow characterizing the essential trade-off offered by hybrid schemes: while a large $N_{\text{RF}}$ generally allows an increased design flexibility \cite{6717211}, the overall power consumption is increased due to the additional analog hardware components required as detailed in Sec.\ \ref{sec:EE}. Overall, it can be observed that, while far from the fully digital system EE, hybrid schemes with reduced $N_{\text{RF}}$ are still capable of offering EE gains over those with large $N_{RF}$ for different $\rho$'s, since their reduced power losses are able to compensate for their theoretical performance degradation.

\section{Conclusion}

In this letter we have described the implications of employing ABFN in hybrid precoding systems when practical losses that are commonly ignored, are taken into account. In particular, we have focused on understanding their implications on the data rates and the EE. The results shown in this letter demonstrate that the performance of realistic hybrid schemes a) is highly dependent on their hardware implementation, where there is a clear distinction in the performance between the DFT and fully connected designs, and b) is significantly  diminished when realistic losses are considered.

\section*{Acknowledgements}

The authors would like to thank Dr.\ Paolo Baracca for his valuable comments on the manuscript.

\bibliographystyle{IEEEtran}
\bibliography{references}
\end{document}